# Political Propagation of Social Botnets: Policy Consequences[1]

Shashank Yadav[2]

### ABSTRACT

The 2016 US election was a watershed event where an electoral intervention by an adversarial state made extensive use of networks of software robots and data driven communications which transformed the interference into a goal driven functionality of man-machine collaboration. Reviewing the debates post the debacle, we reflect upon the policy consequences of the use of Social Botnets and understand the impact of their adversarial operation in terms of catalysing institutional decay, growing infrastructural anxieties, increased industry regulations, more vulnerable Individuals and more distorted ideas, and most importantly, the emergence of an unintended constituency in form of the bot agency itself. The article first briefly introduces the nature and evolution of Social Botnets, and then moves over to discussing the policy consequences. For future work, it is important to understand the agency and collective properties of these software robots, in order to design the institutional and socio-technical mechanisms which mitigate the risk of adversarial social engineering using these bots from interfering into democratic processes.

"The notion of a personal computer is really counterintuitive. There is no such thing as a personal computer. Everyone's computer can be used to attack another country."

> *- Lauri Almann, Estonian Permanent Undersecretary of Defence after Botnets temporarily disabled Estonian government in 2007 as it prepared for remote electronic elections via public internet for the first time*

"We knew the 14 million people we needed to win 270. We targeted those in over 1000 different universes with exactly the things that mattered to them. We won exactly where we laid our money."

> *- Brad Parscale, digital director of the 2016 Trump campaign*

---

# INTRODUCTION

In 44 BC, a nobleman named Octavian launched a smear campaign against Mark Antony. Using short, sharp slogans written upon coins, he painted Mark Antony as having been completely corrupted by Cleopatra, proclaiming that this womanising and drinking has rendered him unfit to rule. Octavian had deployed a clever strategy for information distribution, and the scale and speed of his technique had fundamentally changed the effectiveness of an ordinary rumor (Posetti & Matthews, 2018). Coins have had a natural circulation in human societies, so the word spread quickly and Mark Antony and Octavian eventually came face to face in the first civil war of the Roman Republic. Mark Antony had already lost the support of his publics, and eventually he and Cleopatra, both killed themselves and Rome annexed Egypt. Octavian then took over the void left by Mark Antony and became the first Roman emperor, carefully referring to himself not as the king but as the first citizen. He is better known today as Augustus, who arguably first showed how to hack the modern democracy.

As the above example shows, the modern democracy functions in a mediated environment wherein instead of having any direct knowledge, most voters rely on transmission of information through various media to know anything about their representatives as well as about the associated policy issues, being themselves motivated largely by non-rational forces (Achen et al., 2017). The 2016 elections in the US presented a watershed moment where an intervention by an adversarial state used a combination of cybersecurity and disinformation operations enabled by internet bots, and led to massive public debates and deliberations over the potential of artificial intelligence technologies in influencing elections in another country.

Academics studying election interference or practitioners trying to respond to it have always had to grapple with the challenge of lacking shared definitions as well as the problems of timely identification and attribution when it comes to malicious cybersecurity operations (Martin et al., 2019). To resolve this issue we take the definition set out by Dov Levin (Levin, 2019) which suggests that an intervention is "a situation in which one or more sovereign countries intentionally undertakes specific actions to influence an upcoming election in another sovereign country in an overt or covert manner which they believe will favor or hurt one of the sides contesting that election and which incurs, or may incur, significant costs to the intervener(s) or the intervened country". The 2016 US elections not only fit this criteria, but also also present a template for the use of specific technologies in achieving political objectives.

In this report, we review the materials surrounding such information operations in order to explore the policy consequences of the extensive use of adversarial networks of software robots and data driven communications which transform such political interference into a goal driven functionality of man-machine collaboration.

**SOFTWARE ROBOTS & SOCIAL NETWORKS**

A Social Botnet is defined as a network of software robots that "control online social network accounts and mimic the actions of real users" (Boshmaf et al., 2013). The original software robot (Internet Softbot) was developed in 1993 as a fully implemented AI agent (Etzioni & Weld, 1994) whose actuators included ftp, telnet, mail, and numerous file manipulation commands. Its sensors included internet facilities such as archie, gopher, netfind and others.

Designed to incorporate new facilities into its repertoire as they become available, the software robot was not really "intelligent" but it was a great model for a software based AI agent with its basic architecture envisioned to provide:

- An integrated and expressive interface to the internet

- Dynamically choosing which facilities to invoke and in what sequence

- Behavior change in response to transient system conditions

While a network of software robots is also called a botnet, it is very different from the traditional notion of a botnet in the cybersecurity discourses, which is a network of compromised machines. Although the development paradigm and strategy involved is not so different, the key differences emerge from the command, control, and communication (C3) mechanisms.

A traditional internet botnet made use of IRC and HTTP protocols for its C3 requirements. Around 2011, some botnets were discovered which were using DNS protocols to transmit command and control messages (Negash & Che, 2015). The addition of DNS allowed dynamically changing the IP addresses associated with the botnet, making detection and dismantling of the network much more difficult. The situation was further complicated with the advent of Domain Generation Algorithms (Schiavoni et al., 2014).

By this time, the advent of social networks had already brought in the use of social network's messaging system into the botnets' C3 element (FireEye, 2015), as well as the malwares spreading through social media had given the early indicators of the evolution of traditional botnets into social botnets (Tanner et al., 2010). After the 2016 US elections, the impact of botnets mimicking human users in online social networks unequivocally entered the contemporary public policy debates and became a key focus of research. These type of botnets, i.e. the social botnets, were not just able to lead to a network of infected machines, but also needed for their functioning a set of synthetic digital identities for bots to imitate and socialise with human network users. These digital identities could be human or even computer generated, and again, the simultaneous developments in computational generation are further aiding into this development (Adams, 2017).

Researchers studying online community engagements about crisis events have found that while about 10% of accounts in such conversations have been Social Bots, some groups within these communities are composed entirely of Social Bots alone (Nied et al., 2017). This brings to attention the ability of Social Botnets to influence and/or misrepresent popular opinion during critical events.

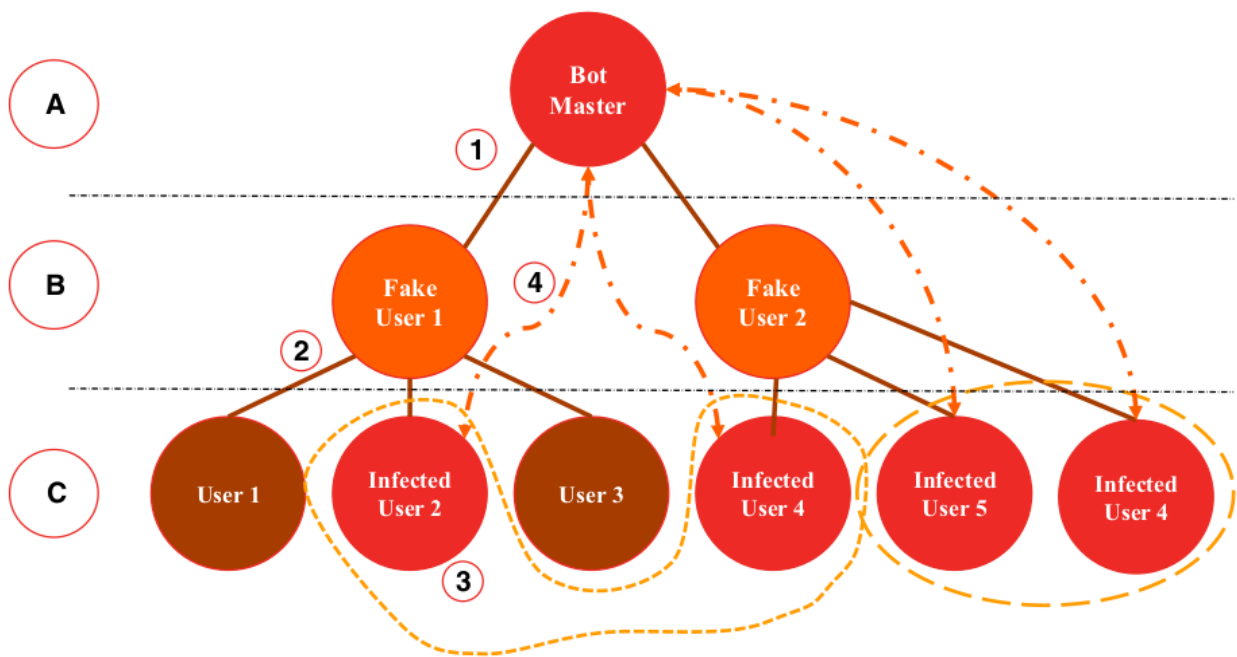

*The propagation of a **Social Botnet**. Image Source: (Faghani & Nguyen, 2019)*

While botnets had long been using social networking sites, exemplified by the KoobFace botnet which not just made a mockery of Facebook in its name, but extensively used the site's messaging system to spread malicious links (Thomas & Nicol, 2010) – the Social Bot brings in the additional property of emulating a human user and using natural language to communicate with other real human users as just another user of the network. As a result of this property, some researchers have gone even further and demonstrated the use of social bots to infiltrate and cultivate specific employees of targeted organisations (Elyashar et al., 2014), showing the extensive breadth of malicious use cases for social botnets.

To put Social Bot detection into perspective, one the earliest large Social Botnets detected emulating humans was discovered by a complete accident in 2017, having survived for four years without being noticed since its creation in 2013 (Echeverría & Zhou, 2017). It consisted of 350,000 live bots randomly tweeting quotes from Star Wars novels – and the researchers who accidentally discovered it during their another experiment with Twitter data, and those who studied it later, could not yet tell if someone was running this just for fun or had any other malicious intent which the size of the botnet might suggest.

Therefore an improvement over the traditional bots, today's social bots not only have the ability to steal and abuse users' identity, but also to impersonate humans and help disseminate a large political campaign in online social networks while evading detection (Zago et al., 2019) as witnessed prominently in the 2016 US elections. A study by the Project on International Peace and Security (Bondy, 2017) described two key strategic logics of deploying social bots in political campaigns:

- *First Mover Advantage*

  It is easier to spread disinformation than to counter it later. Social Botnets offer a stronger first mover advantage in rapidly creating a desired information environment, and reduce the window of opportunity for defenders to introduce any warnings or counter-messaging scheme.

- *Quantity as Quality*

  Ordinary social bots wouldn't pass a thorough Turing test, and they do not have to. A social botnet does not infiltrate a human network by persuasive arguments, but operates on scale and spreads seemingly credible information by making use of the 'majority illusion' effect.

The use of social botnets in Russia's 2016 US campaign has become quite a case study with some scholars noting that from 2016 onwards we have been witnessing the third generation of social bots which are not only much harder to detect than their traditional counterparts, but given the trajectory of technology, will also evolve to become even more harder to detect in future (Cresci, 2019). Simultaneously, large-scale political messaging by social botnets also leads to what researchers have described as "Information Gerrymandering" (Stewart et al., 2019) – as a result of which when a political party uses large Social Botnet based messaging operation for campaigning and wins a disproportionately large share of the votes, other parties are also incentivized to follow the same methods, leaving everyone trapped in a deadlock.

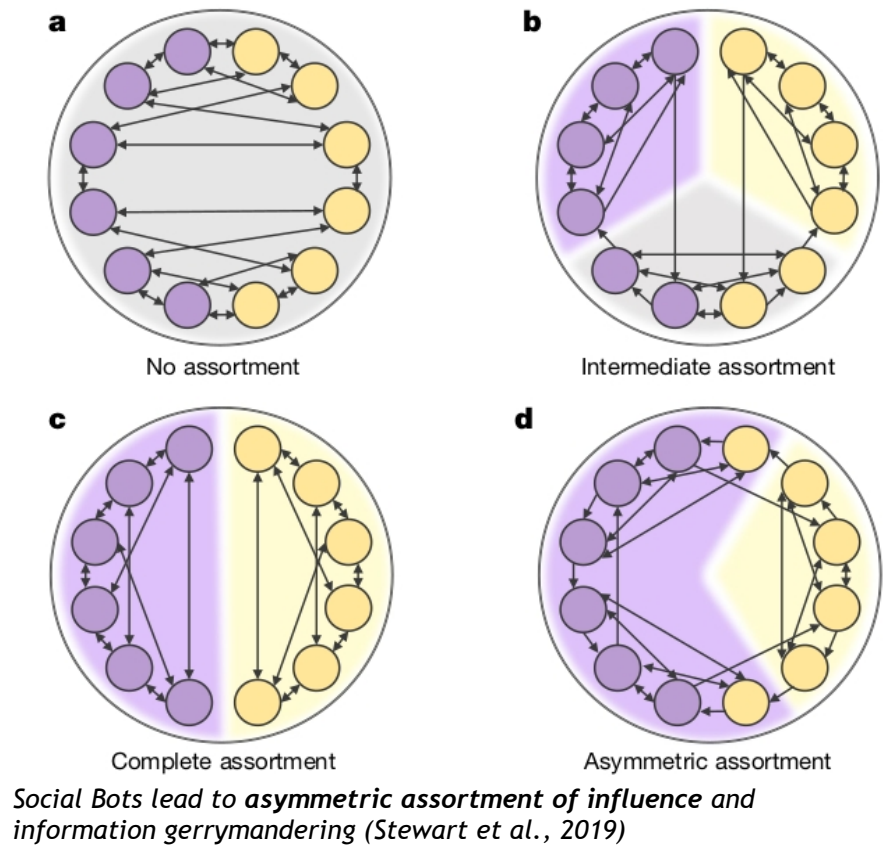

*Social Bots lead to **asymmetric assortment of influence** and information gerrymandering (Stewart et al., 2019)*

Moreover, electoral interventions are more than just informational influence and sometimes also involve these capabilities to compromise physical infrastructure, as in the 2016 US elections where other than conducting partisan electioneering, Russia had also remotely attacked the electronic voting infrastructure of 21 American states (*Russian Interference In The 2016 U.S. Elections*, 2017). In the 2016 US elections, the dominant role that social bots played was highly optimised political messaging (Bessi & Ferrara, 2016), while traditional botnets served the more offensive purposes. It led to the US agencies collaborating with the private sector over a Defending Democracy Program, under which Microsoft took down TrickBot – one of the biggest botnets ever and a prolific distributor of ransomware - before the 2020 US elections, over its perceived threat to disable the electoral and campaign infrastructure (Microsoft, 2020).

Consequently, the social botnet emerges as a multi agent system where an incipient amalgamation of capturing, processing, generating, and delivering of information can be seen within a single agent. These AI capabilities have been understood as the core autonomous capabilities that enable actors to influence, i.e. sense and shape our networked information environment and shift the balance of political stability and security (Yadav, 2022). Since there is a network effect of social media messaging, use of search optimisation, and cross-platform propagation, the infection scope from a Social Botnet is considered to be much larger compared to a traditional botnet (Li et al., 2012), which thus presents the social bot as a demonstrably robust application for an adversarial AI agent.

**POLICY CONSEQUENCES OF SOCIALLY-AWARE ADVERSARIAL AI**

Scholars have identified five key societal verticals which had to bear the consequences of foreign meddling in 2016 elections (Henschke et al., 2020) - these are institutions, infrastructure, industry, ideas, and individuals. Use of software robots and data intensive campaigns also further affects, amplifies and exacerbates the interference's effects on these verticals. In fact, it can be argued that such use of AI brings to table a new kind of problem set for policy makers, that of a machine agency which at micro-levels is somewhat independent of the agency of its developers, and here too social bots have emerged as a key example (Guilbeault, 2016).

Accordingly, we will also draw the policy consequences of electoral propagation of adversarial social botnets into the following phenomena affecting the contemporary policy landscape and choices:

- Institutional Decay,
- Infrastructural Anxieties,
- Industry Regulations,
- Vulnerable Individuals,
- Distorted Ideas, and
- Unintended Constituency

As follows, the 2016 Russian operation into the US electoral ecosystem presents examples from each of the above, further highlighting the difficulty and the utmost urgency of mitigating the strategic risks from dumb robots of today over the superintelligence of tomorrow.

### 1. Institutional Decay

AI in itself is a centralising force, consistent with Carroll Quigley's assertion that the availability of more offensive power leads to a more intense political organisation of the state (Quigley, 2013). For a period of almost three years, the US' Select Committee on Intelligence carried out hearings both open and closed, interviews, and intelligence reviews, to ascertain the validity, extent and nature of Russia's cyber operations in influencing the US national elections in 2016. The committee ascertained that these operations were aimed at regime change in US to achieve the larger Russian policy objectives of the dissolution of European Union (EU) and the North Atlantic Treaty Organization (NATO) (*Open Hearing On The Intelligence Community's Assessment on Russian Activities and Intentions in the 2016 U.S. Elections*, 2017).

The objective of the intervener therefore may not be to just alter the vote share, manipulate voter turnout, or compromise voting infrastructure, but also to undermine trust in important public institutions or/and cause general societal breakdown. Expectantly, the 2016 cyber operations also deeply disturbed the existing institutional frameworks within the US in following aspects:

- ***Diluting the Decentralised Approach to Governance***

    Decentralisation was a key aspect of how US managed its election security before 2016. Operationally, US does not have one national election – it has 50 state elections – and those 50 states had so far managed their own cybersecurity. However post-2016 election, policymakers deemed this setup to be a major vulnerability, as whenever federal cybersecurity assistance was provided, coordination and implementation among various states turned out to be a major problem with each state having its own standards and vendors (*Open Hearing: Election Security*, 2018). Effectively an adversary did not have to defeat one big military grade federal cyber defense but pursue smaller victories at state and local levels, which also added to the criticality of so called "swing states". This decentralisation also meant greater chances of success for adversary while computationally scanning for vulnerabilities. And unlike the cold-war era counter-intelligence networks on ground, responding to computer programs hosted outside national territories also went beyond the limits of local and regional authorities.

    The policy response therefore has been more inclined towards a centralised technical architecture which would make producing a coordinated defense a lot easier, but at the expense of regional autonomy. In fact many academics studying policy frameworks surrounding the 2016 electoral intervention have also recommended the urgency for a centralised or federal body (Ohlin, 2021)

and intra-agency working groups (Henschke et al., 2020) to reduce the risks of existing decentralised institutional framework.

- *Challenging the Existing Norms and Institutions of Deterrence*

    As social bots played a central role in the diffusion of disinformation, spam and malware, and gained much attention following their application in 2016 US elections - a recognition has emerged among among scholars that while timely detection of malicious AI and cyber operations is really hard, it is still easier than prevention and deterrence. Prevention is a hard problem requiring the solution of several socio-technical challenges (Boshmaf et al., 2013).

    However, policymakers aiming for prevention imposed the conventional norms of deterrence which require timely detection and precise attribution, over cyber methods used for an electoral intervention, which mostly lack such detection and attribution. The result has been a contentious policy known as Countering America's Adversaries Through Sanctions Act (CAATSA). A direct response to the Russian meddling in 2016 elections, it provided policymakers with a flexible range of sanctions and financial prohibitions to impose upon countries doing business with the Russian defence sector, with the expressed objective that it will deter adversarial cyber behaviors from Russia. Other legislative attempts to deter election interference included two DETER Acts. Defending Elections from Threats by Establishing Redlines Act was introduced by Marco Rubio and hasn't passed any floor test, and another Defending Elections against Trolls from Enemy Regimes Act was aimed at preventing "inadmissible aliens" who digitally influence US elections from entering the US, which did pass the senate but the matter remains at that. As of this writing, any evidence that such conventional deterrence practices have had the intended effect on malicious digital behaviour, is yet to emerge.

    **2. Infrastructural Anxieties**

Social networks by definition form a good command and control element for social botnets. This draws from the decade old idea that in informatised command and control systems the human element needs to be a part of the system itself and not be placed outside the system boundary (Liu et al., 2011). This is the key to self-synchronous military information operations, the kind that GRU managed to pull off in 2016 in US, wherein an electoral intervention in globalised networked information environments could be described as 'chaotic control'.

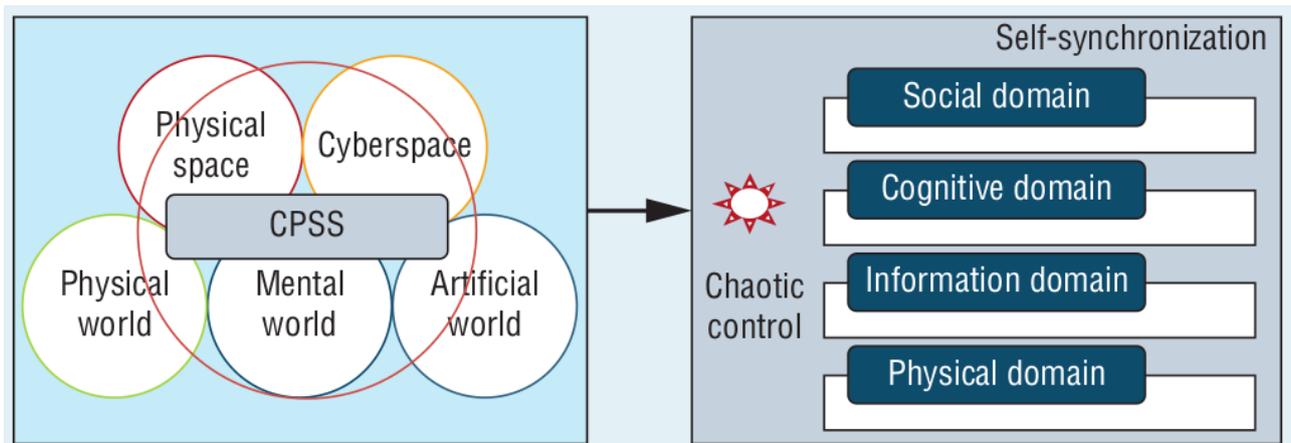

*Command and control logic for state-led information operations generally **integrates the Social**, Cyber, and the Physical worlds. Source (Liu et al., 2011)*

Therefore, combating AI-mediated communications online requires dominance across all information, media and communication infrastructures. A clear consequence of this infrastructurally intensive competition and hostile political engineering is that such capacity, owing to the kind of resources and time required to defend against or orchestrate such operations, requires a combination of technical capability and legal authority which is usually only available with state actors.

Because AI capabilities were used to simulate a false social support which can affect actual voter inclination, the senate intelligence committee hearings show an affirmative inclination towards shifting focus towards greater detection capabilities, increased cyber surveillance and reconnaissance infrastructure, and increased agency budgets (*Open Hearing: Policy Response To The Russian Interference In The 2016 U.S. Elections*, 2018). Admittedly, these operations were remarkably precise and displayed considerable technological savvy, along with a strategic and tactical deftness not seen before. Witnesses and members of the committee have described the campaign as "part-human, part-machine" and a blitzkrieg of information warfare, which eventually led to the designation of elections themselves as a Critical Infrastructure.

Moreover, prominent academics have also echoed these sentiments and advocated further for public defence and publicly funded cybersecurity of political parties and their campaign teams (Baines & Jones, 2018). Such calls for diversion of public funds to cover the insufficiencies of private and quasi-private infrastructure is also a relatively new phenomena arising from the use of computing and communications technology based attacks on democratic societies. The congress also recognised this by affirming that the weakest links in electoral security are the campaign teams, which for the lack of same level of cybersecurity as government agencies, are much more vulnerable to infiltration and exploitation (*Open Hearing: Election Security,* 2018).

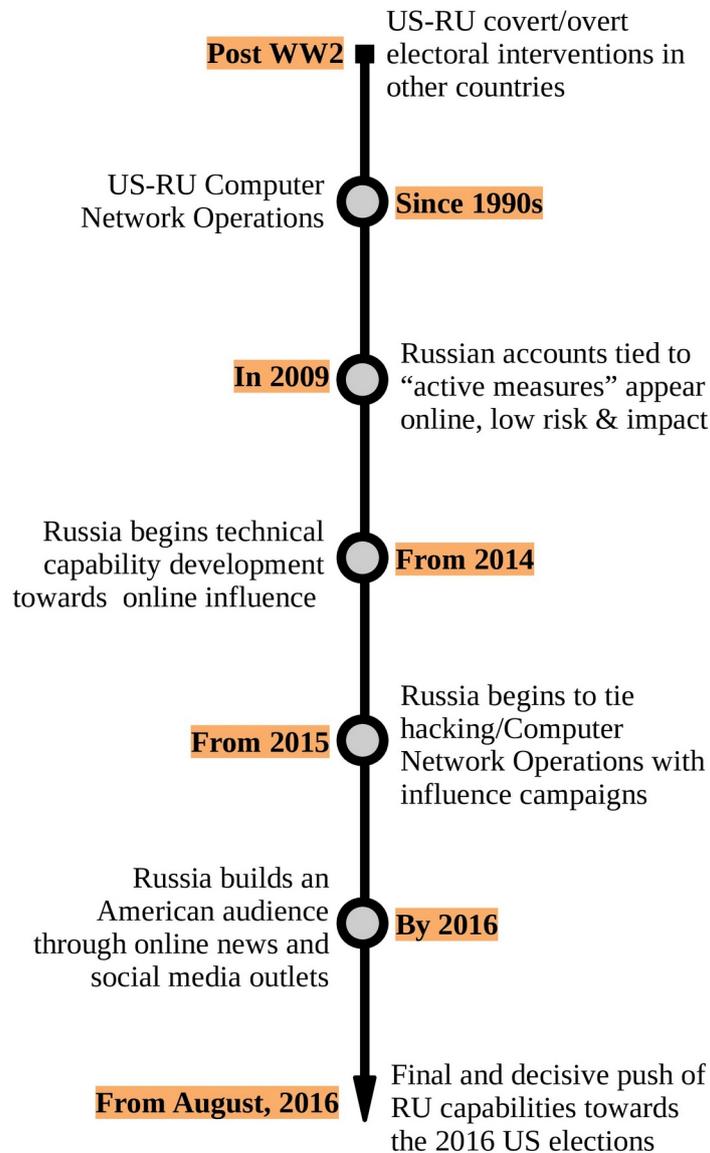

*With prevalent hostile **competition over domains of electronic and network infrastructure**, The 2016 US election was but an inflection point. Image: author*

It must be noted that the infrastructural competition and anxieties that dominated the policy discourse post 2014 intervention also contained an element of shock and awe in the American policymakers at being the receiving end of such interference. Writing for the IEEE computer society, (Berghel, 2017) makes a much needed non-partisan assertion over the policy debates on 2016 interference methods, quipping that merely "the shoes have changed feet". He further suggests that the best short-term response for mitigating ICT enabled electoral interventions would be through technological infrastructure only, i.e. via mobile apps and browser addons etc - and certainly not through any cybersecurity alliance between the principal offenders themselves. This was a direct opposition to some of the policy ideas discussed in the congressional hearing about appointing an international coordinator to salvage US allies' and alliances' electronic and network infrastructure, to respond to malicious cyber behaviors in US networks (*Open Hearing: Policy Response To The Russian Interference In The 2016 U.S. Elections*, 2018).

### 3. Industry Regulations

When it comes to stresses over the functioning of the industries that are directly related to the use of AI in election interference, there emerge three key aspects - unregulated data processing operations, incentivisation of underground technical services, and deeper government-industry collaboration.

- *Unregulated Data Processing Operations*

    Cambridge Analytica's controversial use of inferred data from 50 millions Facebook users through third-party cooperation had been critical to the success of Trump's presidential campaign. The key data analytics and communications company in the 2016 US elections, working at the intersection of international collusions and electoral interventions, went on to win in 2017 a Gold prize in the Big Data category from Advertising Research Foundation (Chester & Montgomery, 2017). Alexander Nix, the CEO of Cambridge Analytica, in fact publicly gave the credit of 2016 electoral swing to his "revolutionary" data-driven communications strategy, even claiming that the results were not only based on the Facebook users' data but also on recent polling data and millions of available voting records over the history of the US presidential elections (Hegazy, 2021). His company had been managing the entire data operation for the Trump campaign since June 2016 (Confessore & Hakim, 2017), and had been at the center of the strategy to invest in more social bots as the communication delivery mechanism, whose numbers had exceeded the opposition campaign's social bots by five to one on the election day (Illing, 2018).

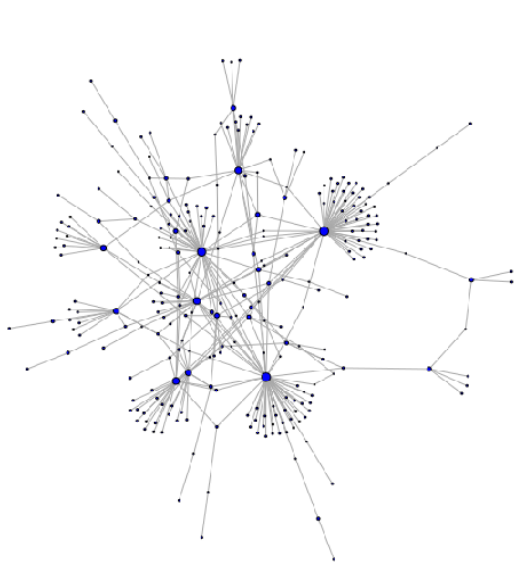
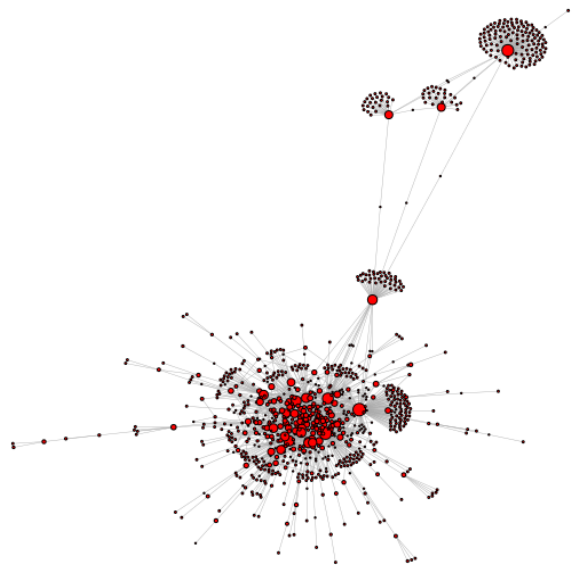

**Clinton Botnet**          **Trump Botnet**

*Social Botnets played an extensive role in propagating computational propaganda during the 2016 US elections and were **a key delivery vehicle** for Cambridge Analytica's data driven political advertising. Image Source: (Woolley & Guilbeault, 2017)*

The senate committee also, as part of Questions for the Record, enquired about the acquisition of fully anonymized, aggregated data from social media companies, while also making a reference to the industry-academia collaboration in AI research (*Open Hearing: Foreign Influence Operations' Use Of Social Media Platforms (Company Witnesses)*, 2018). Interestingly, the analytics company's parent organisation, Strategic Communication Laboratories, served not just the Russian interests but also the US-UK militaries for developing data-driven PsyOps (Bakir, 2020). Notwithstanding, the congress did make a note of the growing trend of outsourcing digital active measures to privateers and the compounding threat of AI driven computational propaganda (*Open Hearing: Foreign Influence Operations' Use Of Social Media Platforms (Third Party Expert Witnesses)*, 2018).

- *Incentivisation of Underground Technical Services*

    Other than any ideological motivations, the operators of fake news operations also had a fairly decent economic incentive, with one operator of a fake site earning close to $16,000 in the final three months of the election campaign during 2016 (Subramanian, 2017). The then US president, Barack Obama, went

on record to point out the "digital gold rush" being experienced by adversarial but legitimate optimisation of search and ad engines (Remnick, 2016).

The use of Social Bots in itself spans a wide and steadily increasing range of activities which include DDoS Attacks, political astroturfing, spreading political adware and malware, cyber espionage, hosting malicious applications and activities on unsuspecting user devices, exploiting SQL vulnerabilities, or even just mining some cryptocyrrency. Thus there is an aspect of profitability too, a clear economic incentive is present for the operator to use these Social Bots to go large scale in infiltrating users' social networks.

Scholars have dubbed this as the "infrastructure of manipulation" which is poised to foster an underground digital economy bustling with hacker-for-hire type of jobs (Frischlich et al., 2020), pointing in their research that just one week's access to a full-service Social Botnet with two live bots (full-service implies that the botnet could also conduct DDoS and other similar cyber activities, over and above social network's narrative manipulation) was sold for 424,35 Euros in a clearnet forum, while a complete full-service botnet itself was offered for 578,99 Euros on a darknet market.

- *Deeper Government-Industry Collaboration*

    The technologically intensive 2016 election interference led to growing calls for building an industry-wide coalition to coordinate and encourage the spread of best practices (McFaul, 2019). The stress for collective response from industry as a foreign policy instrument also laid the foundation for timely information sharing mechanisms between government and industry, and came in the wake of social media companies first having rejected the allegations of electoral abuse of their platforms (*Open Hearing: Policy Response To The Russian Interference In The 2016 U.S. Elections*, 2018).

    Since exploited computers, hacked emails, and persistent Russian presence in US private sector networks had created the central thrust of the 2016 election interference, the strong advocacy for integration of private industry practices into the wider national cybersecurity framework also emerged as a consequence of the electoral intervention.

## 4. Vulnerable Individuals

Rand Waltzman of the RAND Corporation gave a testimony to the Senate Armed Services Committee in April 2017 (Waltzman, 2017), wherein he predicated the electoral vulnerability to foreign digital influence upon two key factors:

- The unprecedented speed and extent of information distribution
- The audience's cognitive vulnerabilities

While the former falls entirely beyond human nature and deals solely with the nature of information and communication technologies, the latter, which is a feature of the human nature, refers to a population being more receptive to certain information as it better appeals to their anxieties and fears. This is also more widely known in advertising industry as neuromarketing, now being heavily utilised in data-driven election campaigns (Hegazy, 2021).

A great and very specific example of this which emerged in the course of 2016 US elections was the use of advanced location-based technologies by Trump campaign to heavily geo-target six crucial states in the final weeks of the elections. These states once used to be the US industrial heartland but had lost many jobs because of automation and offshoring, and some anthropologists believe (González, 2017) that a technologically effective capturing and manipulation of the industrial automation and offshoring related fears and anxieties of populations in these specific geographies was a major factor that led to the election of the 45th US President.

Rand Waltzman also noted in his testimony that the contemporary cybersecurity discourse is primarily concerned with navigating technical features, and little attention is paid to addressing the psychosocial effects of operations on individuals, which is the core component of electoral manipulation by social botnets.

The 2016 US elections effectively demonstrate the utility of socially-aware software systems for adversarial states to influence human behavior, and consequently the national political processes. Inadvertently, this opens up the cognitive-behavioral surface of the electorate itself as a potential target, and not just the weaknesses in electronic voting infrastructure. Therefore, the length of the election season also expands the electoral attack surface from cyber enabled digital manipulation techniques. Intuition and available evidence both seem to support this hypothesis (Hansen & Lim, 2019), that longer election season leads to greater socio-technical vulnerabilities being exposed for longer duration.

The temporal dimension is very important for cognitive-behavioral aspects of a bot led intervention, as some researchers also discovered that exploiting human traits such as reciprocity and confirmation bias, the misinformation carrying communication had highest spread potential around the election night itself (Oehmichen et al., 2019).

## 5. Distorted Ideas

AI-mediated communication has been defined as "mediated communication between people in which a computational agent operates on behalf of a communicator by modifying, augmenting, or generating messages to accomplish communication or interpersonal goals"(Hancock et al., 2020). Furthermore, tools like social botnets have been proven to be really effective for rapidly generating small to medium-sized information cascades in critical elections (Bastos & Mercea, 2019).

This is very pertinent because the consumption of particular type of news has defining consequences for voter turnout and real-world political participation (DellaVigna & Kaplan, 2007). In weeks leading up to the 2016 US elections, the electorate's engagement with the "fake news" stories had exceedingly increased in relation to their engagement with the news from mainstream outlets (Budak, 2019), creating a distorted information environment.

Going into further depth over political manipulation, another set of researchers (Grimme et al., 2017) suggested three key challenges from an adversary's perspective, which being rooted in deception, further highlight the creation of conspiratorial and distorted information environment as the defining feature of large-scale Social Botnet led information operations:

- Producing credible and intelligent content
- Leaving a trace of human like metadata
- Cultivating an adequate and balanced online network

Linvill et al. (2019) contend that the purpose of state sponsored distribution of (dis)information, particularly in another state's electoral ecosystem, is to drive the process of agenda building. This allows the attacker to shift the focus away from policymakers and towards the polity itself. Russia used its carefully cultivated technical capabilities in the same manner, to promote the acceptability of its own policy objectives within the US population and consequently subvert the polity.

The automation and distortion of narratives has been described as the key characteristics of the landscape of contemporary political communication where researchers find a clear connection between bots, hyper-partisan media outlets, and political fringes (Ferrara et al., 2020). Since social bots are developed to fully automate the behavior of a social media account, the better a social bot is, the harder it is to detect. This is quite akin to the generative component of a GAN being used to create synthetic information, where greater generative accuracy besets the classification of falsehoods. And unlike humans, who generally propagate their own political leanings, bots tend to propagate ideas from all political sides – catalysing self-reinforcing narratives across the entire political spectrum.

Robert Mueller's indictment on Internet Research Agency also went further on the effects of social botnet operations, asserting that Americans were driven to attend real-world protests as a result of Russian campaign disinformation online (Mueller III, 2020). For example, using two separate Facebook pages with bot activity simulating social support, two protest groups opposed to each other were brought at the same spot in front of the Islamic Da'wah Centre of Houston to protest each other at same time (Hanson et al., 2019).

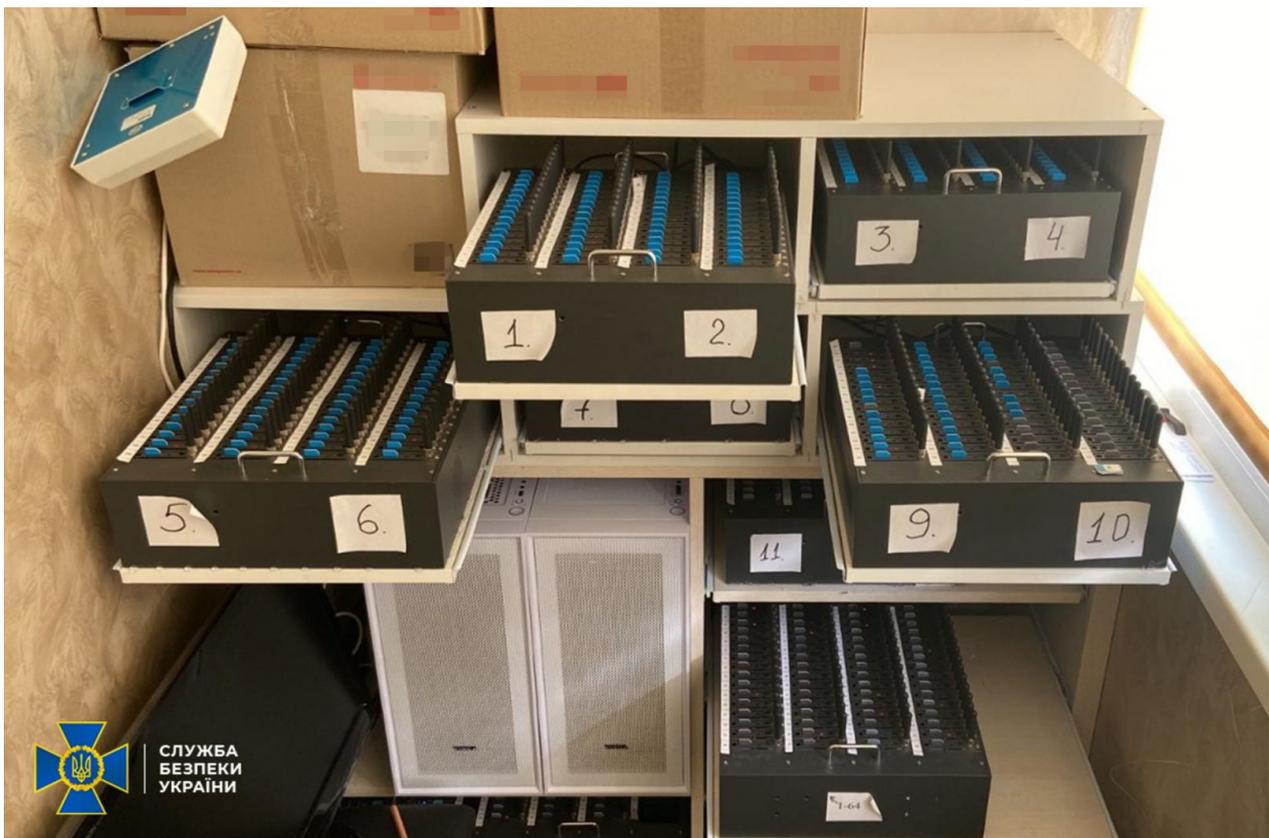

*A Look Under the Hood: Here are terminals with SIM cards **emulating unique and real human users** in the virtual social world, this particular social botnet consisted of 100 sets of GSM gateways, over 10000 SIM cards, and laptops/computers for coordination and control. Distorting conflict situation and demoralising its adversary, it was recently captured by Ukraine post the Russian special military operation began in 2022 (Toulas, 2022).*

## 5. Unintended Constituency

Growth in AI capabilities has led societies into an environment where humans are not the only agents capable of persuasion, online social network as an environment propels bots into their own agency (Guilbeault, 2016). The internet thus has come to be a new kind of habitat and bots have emerged as its very own indigenous species. Experiments suggest that social bots need only constitute 5%–10% of participants in an online discussion for the view being propagated by them to eventually become the dominant opinion held by over 2/3 of users in that discussion (Cheng et al., 2020).

In one study over the impact of software robots on US 2016 elections, scholars found that close to one fifth of the entire political conversation was computer generated (Bessi & Ferrara, 2016). They outlined three further complications which emerge from this new constituency of AI programs:

- Redistribution of influence with malicious intent
- Further polarisation of political conversations
- Enhanced distribution of unverified and low credibility information

Most importantly, large populations of social bots exhibit properties of collective behavior such as those found in large socio-biological systems (Duh et al., 2018). The effects that such bots have on public opinion and their ability to swing vote share in critical elections inevitably makes them an important political actor. Since digital platforms serve as a fundamental infrastructure for political conversations today, automated political communication mediated by increasingly intelligent bots continues to shape the political culture which we live in, even leading to the "spiral of silence" with aggressive bots reducing the human participants' willingness to share their political opinions (Cheng et al., 2020).

Experts have also noted the disproportionate amount of automated bots in developing an unwitting population during the 2016 elections (Shao et al., 2018), particularly through the Trump campaign where Trump's social bots had overwhelmingly outnumbered Hillary's social bots (Illing, 2018). While such socially-aware software systems can exploit the social behavior of an electorate, according to Boshmaf et al. (2013) the successful and sustainable design of a social botnet also requires it to hide its true nature. Therefore, Social Botnets also incorporate internal heuristics for large scale infiltration while hiding their botmaster. These deceptive design elements illustrate an information asymmetry in online AI-mediated communications and increase the security and political risks from fully or partially automated, goal driven malicious autonomous software in social networks.

## CONCLUDING REMARKS

The risks of any particular threat are not assessed in terms of the actual damage it has caused so far, but by the potential damage and disruption it represents into the future. Theoretically, a social botnet could be developed in a certain country, be purchased by a state or non-state anonymous entity from another country, deploying digital identities attributing malicious behaviour to yet some other country or group, while utilising infected machines from countries across the world, to conduct an attack on elections or other key democratic institutions and organisations in a different state's territory altogether. This sets the nature of the puzzle of mitigation, wherein the architecture of response must emerge globally but converge locally.

In retrospect, it is a tribute to Turing that "bot or not" is becoming one of the defining security questions of our age. In order to deter malicious cyber behavior as seen in 2016 US elections, contentious policy acts like CAATSA came to be, but the intended efficacy of such measures remains mired in doubt. Notwithstanding, the most powerful deployments of computational propaganda involve both algorithmic distribution as well as human curation. With the help of their human handlers, these software robots propagate low credibility content, infiltrate into human communities, amplify and simulate social support, influence the indicators of public opinion, and enable a range of other malicious cyber capabilities generally associated with traditional botnets. Furthermore, their use and ease of development leads to the development of a burgeoning underground digital economy and worldwide hacker-for-hire ecosystems.

Given the increasing difficulty in social bot detection, it is therefore of utmost importance to design institutional and socio-technical mechanisms that mitigate the risk of adversarial social engineering using networks of software robots from interfering into democratic processes.

***